\documentclass[
    ,final            % use final for the camera ready runs
  ]
  {aipproc}

\layoutstyle{8x11single}

\begin{document}

\title{Bistability: a common feature in some `aggregates' of logistic maps}


\author{Ricardo López-Ruiz}
{
  address={DIIS and BIFI, Facultad de Ciencias, Universidad de Zaragoza,
  E-50009 Zaragoza, Spain.}
}

\author{Danièle Fournier-Prunaret}
{
  address={SYD and LESIA,
  Institut National des Sciences Appliquées,
  31077 Toulouse Cedex, France.}
}

\begin{abstract}
 As it was argued by Anderson [Science 177, 393 (1972)], the `reductionist' 
 hypothesis does not by any means imply a `constructionist' one. Hence, 
 in general, the behavior of large and complex aggregates of elementary 
 components can not be understood nor extrapolated from the properties of 
 a few components. Following this insight, we have simulated different 
 `aggregates' of logistic maps according to a particular coupling scheme. 
 All these aggregates show a similar pattern of dynamical properties, 
 concretely a bistable behavior, that is also found in a network of many
 units of the same type, independently of the 
 number of components and of the interconnection topology. 
 A qualitative relationship with brain-like systems is suggested.
\end{abstract}

\maketitle

%%%%%%%%%%%%%%%%%%%%%%%%%%%%%%%%%%%%%%%%%%%%
%% MAINMATTER
%%%%%%%%%%%%%%%%%%%%%%%%%%%%%%%%%%%%%%%%%%%%

\section{Introduction}

One of the most challenging scientific problems today is to understand
how the millions of neurons of our brain give rise to the emergent
property of thinking \cite{sfn}. Different aspects of neurocomputation take
contact on this problem: how brain stores information and how brain
processes it to take decisions or to create new information.
Other universal properties of this system are more evident.  One of
them is the existence of a regular daily behavior: the sleep-wake
cycle \cite{winfree, bar-yam}. The internal circadian rhythm is closely synchronized with the
cycle of sun light. Roughly speaking and depending on the particular
species, the brain is awake during the day and it is slept during the
night, or vice versa.
All mammals and birds sleep. There is no a well established law relating
the size of the animal with the daily time it spends sleeping, but, in general, 
large animals tend to sleep less than small animals.
Hence, at first sight, the bistable sleep-wake behavior seems not depend 
on the precise architecture of the brain nor on its size. If we represent the brain 
as a complex network this property would mean that this possible bistability should not 
depend on the topology (structure) nor on the number of nodes (size) of the network.

So, on one side, it has been recently argued in \cite{eguiluz} that the distribution 
of functional connections $p(k)$ in the human brain,
where $p(k)$ represents the probability of finding 
an element with $k$ connections to other elements of the network,
follows the same distribution of a scale-free network \cite{barabasi}. 
Thus, in that work \cite{eguiluz}, the human brain is divided
in $34x64x64$ sites (called {\it voxels}) 
and the magnetic resonance activity of all the voxels 
is recorded. Let us observe at this point that if the human 
brain has around $10^{11}$ neurons, then a voxel has around $10^5$ neurons. 
The calculation of the correlation matrix among the set of the voxels 
activities shows a power law behavior, $p(k)\sim k^{-\gamma}$, 
with $\gamma$ around $2$. This finding means that there are regions in the
brain that participate in a large number of tasks while most of the regions
are only involved in a tiny fraction of the brain's activities.

On the other side, it has been shown by Kuhn et al. \cite{kuhn} 
the nonlinear processing of synaptic inputs in cortical neurons.
They studied the response of a model neuron with a simultaneous increase of 
excitation and inhibition. They found that the firing rate 
of the model neuron first increases, reaches a maximum, and then decreases at 
higher input rates. Functionally, this means that the firing rate, commonly
assumed to be the carrier of information in the brain, is a non-monotonic
function of balanced input. These findings do not depend on details of the model and, 
hence, are relevant to cells of other cortical areas as well. 

\begin{figure}[h]
 \includegraphics[height=.1\textheight]{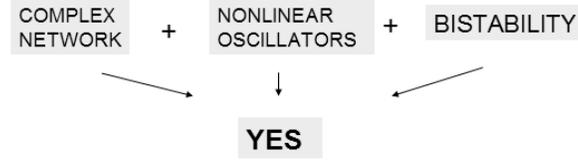}
  \caption{QUESTION: Is it possible to implement some kind of coupling and 
  nonlinear dynamics in each node of a complex network in order to get bistability?. 
  ANSWER: Yes.}
  \label{yes}
\end{figure}

Putting together all these facts, we arrive to the central question that we want
to bring to the reader: are we able to reproduce the bistability in a complex
network independently of the topology and of the number of nodes?. The answer
is 'yes' (Fig. \ref{yes}). What kind of local dynamics and coupling among nodes must be implemented
in order to get this behavior?. In the next section, 
we give a possible strategy for the coupling and the local dynamics 
which should be implemented in a few or many units network in order to find bistable behavior.
An example of this implementation is given in \cite{lopezruiz07}.
Here we center our attention in the bistability present in the case of 
a few coupled functional units. Four different discrete models in two and three dimensions
are collected. In view of the results, we want to suggest with all these examples the possibility 
of `constructivism' in the world of complex systems.

\section{Models of a few coupled functional units}

\subsection{General model}

Our approach consider the so called 
{\it functional unit}, i.e. a neuron or group of neurons (voxels), as a discrete 
nonlinear oscillator with two possible states: active (meaning one type of activity) 
or not (meaning other type of activity).
Hence, in this naive vision of the brain as a networked system, 
if $x_n^i$, with $0<x_n^i<1$, represents a measurement of the $ith$ 
functional unit activity at time $n$, 
it can be reasonable to take the most elemental local nonlinearity, 
for instance, a logistic evolution \cite{may}, which presents a quadratic term, 
as a first toy-model for the local neuronal activity:
\begin{equation}
x^i_{n+1} = \bar p_i\;x^i_n(1-x^i_n).
\label{eq0}
\end{equation}

\begin{figure}[h]
 \includegraphics[height=.05\textheight]{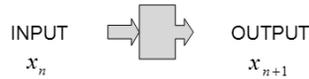}
  \caption{Discrete nonlinear model for the local evolution of a functional unit.}
  \label{1-osc}
\end{figure}

It presents only one stable state for each $\bar p_i$. Then, there is no
bistability in the basic component of our models. For $\bar p_i<1$,
the dynamics dissipates to zero, $x_n^i=0$, then it can represent the
functional unit with no activity. For $1<\bar p_i<4$, the dynamics is
non null and it would represent an active functional unit. This local
transition is controlled by the parameter $\bar p_i$.  The functional
dependence of this local coupling on the neighbor states is essential
in order to get a good brain-like behavior (i.e., as far as the
bistability of the sleep-wake cycle is concerned) of the network. 
As a first approach, we can take $\bar p_i$ as a linear 
function depending on the actual mean value, $X_n^i$, of the neighboring 
signal activity and expanding the interval $(1,4)$ in the form:
\begin{eqnarray}
\bar p_i & = & p_i\;(3X_n^i+1), \;\;\;\;\;\;\; (excitation \;\; coupling) \label{eq1}\\
& {or} & \nonumber \\
\bar p_i & = & p_i\;(-3X_n^i+4), \;\;\;\; (inhibition \;\; coupling) \label{eq11}
\end{eqnarray}
with 
\begin{equation}
X_n^i={1\over N_i}\sum_{j=1}^{N_i}x_n^j.
\label{eq2}
\end{equation}
$N_i$ is the number of neighbors of the $ith$ functional unit, and
$p_i$, which gives us an idea of the interaction of the functional unit
with its first-neighbor functional units, is the control parameter.
This parameter runs in the range $0<p_i<p_{max}$, where $p_{max}\succeq
1$. When $p_i=p$ for all $i$, the dynamical behavior of these networks 
with the excitation type coupling \cite{lopezruiz07} presents an attractive global
null configuration that has been identified as the {\it turned off}
state of the network.  Also they show a completely synchronized
non-null stable configuration that represents the {\it turned on}
state of the network. Moreover, a robust bistability between these two perfect 
synchronized states is found in that particular model 
(see \cite{lopezruiz07} for more details). For different models with a few
coupled functional units we sketch in the next subsections the regions 
where they present a bistable behavior. The details of the complete unfolding
of these dynamical systems can be found in the 
references \cite{lopezruiz04,lopezruiz05,fournier06}.

\subsection{Models of two functional units}

Let us start with the simplest case of two interconnected $(x_n,y_n)$
functional units. Three different combinations of couplings are possible:
$(excitation, excitation)$, $(excitation, inhibition)$ and 
$(inhibition, inhibition)$. 

\begin{figure}[h]
 \includegraphics[height=.05\textheight]{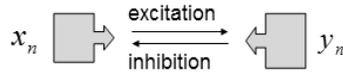}
  \caption{Two functional coupled units.}
\end{figure}

The first two cases of coupling present bistability in different regions of the 
parameter space. The third case also shows bistability in a very narrow
interval of parameter space but it requires a very fine inspection \cite{mira}
that we shelve for a further work.

\subsubsection{Model with mutual excitation}

\begin{figure}[h]
 \includegraphics[height=.2\textheight]{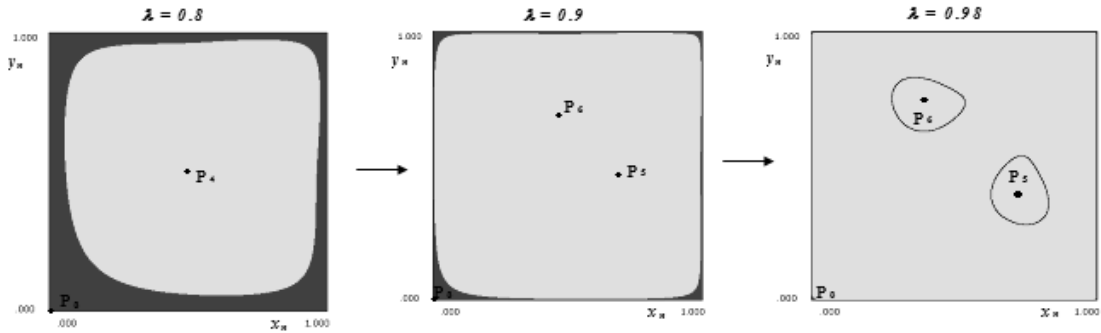}
  \caption{Bistability in $2$ functional units with excitation type coupling.}
  \label{2func+}
\end{figure}

The dynamics of the $(excitation,excitation)$ case \cite{lopezruiz04}
is given by the coupled equations:
\begin{eqnarray}
x_{n+1} & = & p \;(3y_n+1)x_n(1-x_n),\\
y_{n+1} & = & p \;(3x_n+1)y_n(1-y_n).
\label{2-osc}
\end{eqnarray}
The regions of the parameter space (Fig. \ref{2func+})
where we can find bistability are:

\begin{itemize}
\item For $0.75<p<0.86$, the synchronized state, $x_+=(\bar x, \bar x)=P_4$,
with $\bar x={1\over 3}\{1+(4-{3\over p})^{1\over 2}\}$, which arises
from a saddle-node bifurcation for the critical value $p=0.75$, is a
stable {\it turned on} state.  This state coexists with the 
{\it turned off} state $x_\theta=0$.
The system presents now bistability and depending on the initial
conditions, the final state can be $x_\theta$ or $x_+$. Switching on
the system from $x_\theta$ requires a level of noise in both functional units
sufficient to render the activity on the basin of attraction of $x_+$.
On the contrary, switching off the two functional units network can be done, for
instance, by making zero the activity of one functional unit, or by doing the
coupling $p$ lower than $0.75$.
\item For $0.86<p<0.95$, the active state of the network is now a
period-$2$ oscillation, namely the period-$2$ cycle $(P_5,P_6)$ in Fig. \ref{2func+}. 
This new dynamical state bifurcates from $x_+$ for $p=0.86$.  
A smaller noise is necessary to activate the system
from $x_\theta$.  Making zero the activity of one functional unit continues to
be a good strategy to turn off the network.
\item For $0.95<p<1$, the active state acquires a new frequency and
presents quasiperiodicity (the invariant closed curves of Fig. \ref{2func+}).  
It is still possible to switch off the
network by putting to zero one of the functional units.
\end{itemize}

\subsubsection{Model with excitation + inhibition}

\begin{figure}[h]
 \includegraphics[height=.2\textheight]{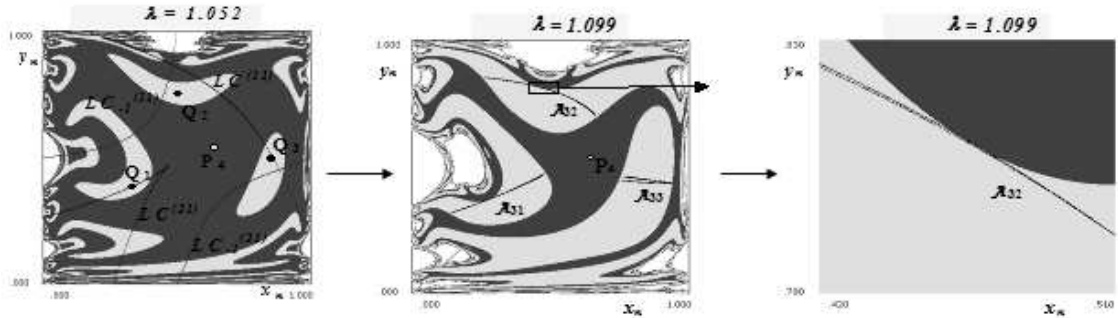}
  \caption{Bistability in $2$ functional units with excitation+inhibition type coupling.}
  \label{2func-}
\end{figure}

The dynamics of the $(excitation,inhibition)$ case \cite{lopezruiz05}
is given by the coupled equations:
\begin{eqnarray}
x_{n+1} & = & p\;(3y_n+1)x_n(1-x_n),\\
y_{n+1} & = & p\;(-3x_n+4)y_n(1-y_n).
\label{2-oscc}
\end{eqnarray}
The regions of the parameter space (Fig. \ref{2func-})
where we can find bistability are:

\begin{itemize}
\item For $1.051<p<1.0851$, a stable period three cycle $(Q_1,Q_2,Q_3)$ appears 
in the system. It  coexists with the fixed point $P_4$. When $p$ is increased, 
a period-doubling cascade takes place and generates successive cycles of higher 
periods $3x2^n$. The system presents bistability. Depending on the initial conditions, 
both populations $(x_n,y_n)$ 
oscillate in a periodic orbit or, alternatively, settle down in the fixed point.
The borders between the two basins are complex. 
\item For $1.0851<p<1.0997$, an aperiodic dynamics is possible. 
The period-doubling cascade has finally given birth to an order three cyclic chaotic 
band(s) $(A_{31},A_{32},A_{33})$. The system can now present an irregular oscillation 
besides the stable equilibrium with final fixed populations. The two basins are now fractal.
\end{itemize}

\subsection{Models of three functional units}

Following the strategy given by relation (\ref{eq1}-\ref{eq11}) several models with 
three functional units can be established. We have studied in detail two of 
them \cite{fournier06} and their bistable behavior is reported here.

\subsubsection{Model with local mutual excitation}

Let us start with the case of three alternatively interconnected $(x_n,y_n,z_n)$
functional units under a mutual excitation scheme. 

\begin{figure}[h]
 \includegraphics[height=.1\textheight]{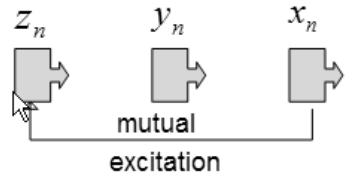}
  \caption{Three alternatively coupled functional units under the excitation scheme.}
\end{figure}

Then the dynamics of the system is given by the coupled equations:
\begin{eqnarray}
x_{n+1} & = & p \;(3y_n+1)x_n(1-x_n), \\
y_{n+1} & = & p \;(3z_n+1)y_n(1-y_n), \\
z_{n+1} & = & p \;(3x_n+1)z_n(1-z_n).
\label{3-osc+}
\end{eqnarray}

\begin{figure}[h]
 \includegraphics[height=.2\textheight]{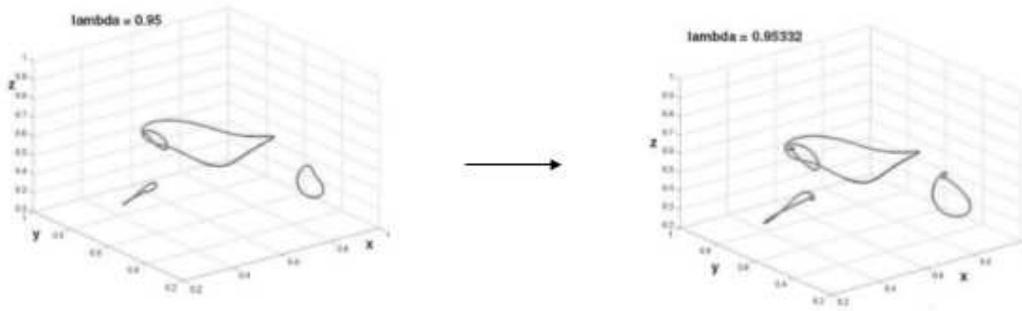}
  \caption{Bistability in $3$ functional units with local excitation type coupling.}
  \label{3func+}
\end{figure}

The regions of the parameter space 
where we have found bistability are:

\begin{itemize}
\item For $0.93310<p<0.95334$, a big invariant closed curve (ICC) $C1$ coexists 
with a period-$3$ orbit that bifurcates, first to an order 
$3$-cyclic ICC (Fig. \ref{3func+}), and finally to an order-$3$ 
weakly chaotic ring (WCR) before disappearing.
\item For $0.98418<p<0.98763$, the ICC $C1$ coexists with another ICC $C2$ 
(see Ref. \cite{fournier06}) that becomes chaotic, by following a period doubling
cascade of tori, before disappearing. 
\item For $1.00360<p<1.00402$, the ICC $C1$ coexists with a high period orbit
that gives rise to an ICC $C3$. This ICC also becomes a chaotic band
(see Ref. \cite{fournier06}) by following a period doubling
cascade of tori before disappearing. 
\end{itemize}

\subsubsection{Model with global mutual excitation}

We expose now the case of three globally interconnected $(x_n,y_n,z_n)$
functional units under a mutual excitation scheme. 

\begin{figure}[h]
 \includegraphics[height=.1\textheight]{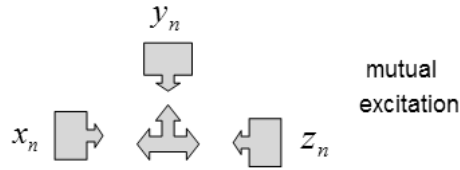}
  \caption{Three globally coupled functional units under the excitation scheme.}
\end{figure}

Then the dynamics of the system is given by the coupled equations:
\begin{eqnarray}
x_{n+1} & = & p\; (x_n+y_n+z_n+1)x_n(1-x_n), \\
y_{n+1} & = & p\; (x_n+y_n+z_n+1)y_n(1-y_n), \\
z_{n+1} & = & p\; (x_n+y_n+z_n+1)z_n(1-z_n).
\label{3-osc-}
\end{eqnarray}

\begin{figure}[h]
 \includegraphics[height=.2\textheight]{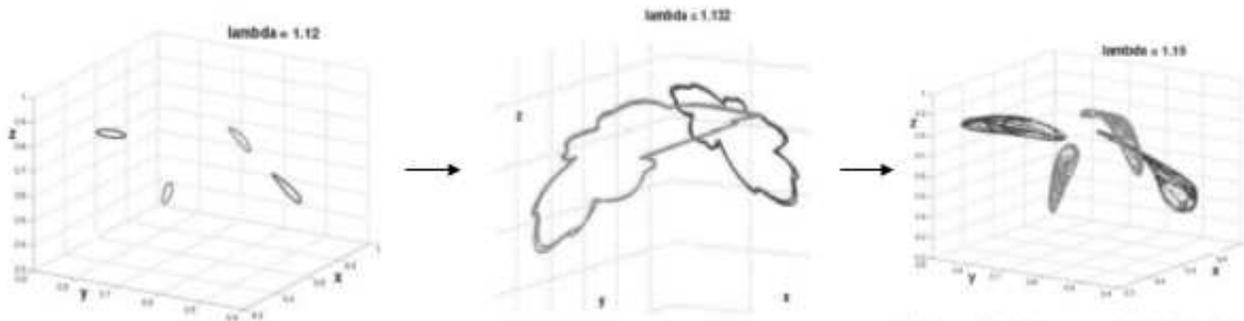}
  \caption{Bistability in $3$ functional units with global excitation type coupling.}
  \label{}
\end{figure}

For the whole range of the parameter, $0<p<1.17$, bistability is present in this system:

\begin{itemize}
\item   Firstly, two order-$2$ cyclic ICC coexist before becoming two order-$2$ 
cyclic chaotic attractors by contact bifurcations of heteroclinic type.
Finally the two chaotic attractors become a single one before disappearing.
\end{itemize}

\section{Conclusions}

One of the more challenging problems in nonlinear science is the goal 
of understanding the properties of neuronal circuits \cite{varona}. 
Synchrony and bistability are two important dynamical behaviors found
in those circuits.  
In this work, different coupling schemes for networks with local logistic
dynamics are proposed. It is observed that these types of couplings
generate a global bistability between two different dynamical states.
This property seems to be topology and size independent. This is a direct
consequence of the local mean-field multiplicative coupling among the
first-neighbors. If a formal and naive relationship is established between 
these two states and the sleep-wake states of a brain, respectively, 
one would be tempted to assert that these types of couplings in a network,
regardless of its simplicity, give us a good qualitative model for
explaining that specific bistability. 
Following this insight, different low-dimensional systems with logistic components  
coupled under these schemes have been presented. The regions where the dynamics
shows bistability have been identified. Other low and high
dimensional models merit a similar detailed inspection in the future. 
This study could put in evidence the possibility 
of `constructivism' in the world of complex systems.

%%%%%%%%%%%%%%%%%%%%%%%%%%%%%%%%%%%%%%%%%%%%
%% BACKMATTER
%%%%%%%%%%%%%%%%%%%%%%%%%%%%%%%%%%%%%%%%%%%%

\begin{theacknowledgments}
R. L.-R. acknowledges some financial support of the 
Spanish DGICYT Projects FIS2005-06237 and FIS2006-12781-C02-01.
\end{theacknowledgments}

\end{document}